\documentclass[%
reprint,
superscriptaddress,
amsmath,amssymb,
aps,
prb,
floatfix,
]{revtex4-1}
\usepackage[version=3]{mhchem}
\usepackage{graphicx}
\usepackage{dcolumn}
\usepackage{bm}

\usepackage[utf8]{inputenc}
\usepackage[T1]{fontenc}
\usepackage{mathptmx}
\usepackage{bbold}
\usepackage{float}
\usepackage{booktabs}
\usepackage{multirow}
\usepackage{amsmath}
\usepackage{amssymb}
\usepackage{mathtools}
\usepackage{xcolor}

\begin{document}

\title{Assessment of the \textit{Ab Initio} Bethe-Salpeter Equation Approach for the Low-Lying Excitation Energies of Bacteriochlorophylls and Chlorophylls}
\author{Zohreh Hashemi}
\affiliation{Institute of Physics, University of Bayreuth, Bayreuth 95440, Germany}
\author{Linn Leppert}
\email{l.leppert@utwente.nl}
\affiliation{MESA+ Institute for Nanotechnology, University of Twente, 7500 AE Enschede, The Netherlands}
\affiliation{Institute of Physics, University of Bayreuth, Bayreuth 95440, Germany}

\begin{abstract}
Bacteriochlorophyll and Chlorophyll molecules are crucial building blocks of the photosynthetic apparatus in bacteria, algae and plants. Embedded in transmembrane protein complexes, they are responsible for the primary processes of photosynthesis: excitation energy and charge transfer. Here, we use \textit{ab initio} many body perturbation theory within the $GW$ approximation and Bethe-Salpeter equation (BSE) approach to calculate the electronic structure and optical excitations of Bacteriochlorophylls \textit{a}, \textit{b}, \textit{c}, \textit{d} and \textit{e} and Chlorophylls \textit{a} and \textit{b}. We systematically study the effects of structure, basis set size, partial self-consistency in $GW$, and the underlying exchange-correlation approximation, and compare our calculations with results from time-dependent density functional theory, multireference RASPT2 and experimental literature results. We find that optical excitations calculated with $GW$+BSE are in excellent agreement with experimental data, with an average deviation of less than 100\,meV for the first three bright excitations of the entire family of (Bacterio)chlorophylls. Contrary to state-of-the-art TDDFT with an optimally-tuned range-separated hybrid functional, this accuracy is achieved in a parameter-free approach. Moreover, $GW$+BSE predicts the energy differences between the low-energy excitations correctly, and eliminates spurious charge transfer states that TDDFT with (semi)local approximations is known to produce. Our study provides accurate reference results and highlights the potential of the $GW$+BSE approach for the simulation of larger pigment complexes.

%\section{TOC Graphic}
%\begin{figure}
% 	    \includegraphics{TOC.pdf}
%\end{figure}
\end{abstract}

\maketitle
\section{Introduction}
Electronic excitations form the foundation of some of the most fundamental natural processes. In photosynthesis, plants, algae and bacteria convert solar energy into chemical energy, utilizing a cascade of coupled energy and charge transfer excitations that are performed by pigment-protein complexes with high quantum efficiency. Bacteriochlorophyll (BCL) and Chlorophyll (CL) molecules are among the most important building blocks of these pigment-protein complexes \cite{Blankenship2014}. They are responsible for the absorption and transfer of excitation energy, and for the charge separation necessary for establishing a proton gradient that eventually drives the synthesis of chemical energy in plants and bacteria \cite{Hu2002}. Accurately calculating the electronic structure and excitations of these molecules from first principles, is the prerequisite for understanding their interactions with each other and with the surrounding proteins and, consequently, energy and charge transfer in natural photosynthesis.

BCL and CL molecules constitute a family of substituted tetrapyrroles with varying absorption properties depending on conjugation and the number and nature of substitutions. CL~\textit{a} and \textit{b} are present in plants and green algae, whereas green bacteria mostly rely on BCL~\textit{c}, \textit{d} and \textit{e} for excitation energy transfer, and BCL~\textit{a} for concentrating excitations close to the reaction center of the photosynthetic unit \cite{Croce2014}. BCL~\textit{a} is also the main pigment in purple bacteria, whose light harvesting apparatus and reaction center are among the most thoroughly studied natural light-harvesting systems \cite{Cogdell2006}. The optical excitation spectrum of these pigments possesses two characteristic absorption bands: 1.~The $Q$ band in the visible part of the spectrum, comprised of excitations $Q_y$ and $Q_x$ with high and low oscillator strength, respectively, 2.~The $B$ (or Soret) band in the near ultraviolet. 

In the field of finite organic and biological molecular systems, neutral excitations and optical spectra are predominantly calculated using time-dependent density functional theory (TDDFT). In conjunction with model Hamiltonian approaches, TDDFT has been employed for the simulation of large photosynthetic pigment-protein complexes \cite{Jang2018,Shao2020}. The accuracy of its approximations and implementations has been tested for a variety of biochromophores \cite{Schelter2019,Sundholm2003,Vokacova2007}. However, TDDFT's standard approximations are inadequate for describing long-range charge transfer excitations \cite{Dreuw2004} and high-energy Rydberg states \cite{Tozer1998} due to self-interaction errors and an incorrect asymptotic behavior. And while exchange-correlation (xc) functionals that contain long-range exact exchange, such as optimally-tuned range-separated hybrid functionals (OT-RSH) can be employed as a remedy in such cases \cite{Kuritz2011,Kuemmel2017}, use of such functionals requires a tedious per-system tuning procedure.

Multireference wavefunction-based methods have scarcely been used for molecules as large as BCL and CL. Vertical excitation energies of CL~$a$ based on ADC(2) and different coupled cluster approaches show a spread of $\sim$0.4\,eV, strongly depending on the method, basis set, and structural model used in these calculations \cite{Suomivuori2016, Suomivuori2019, Sirohiwal2020}. In 2016 and 2019, Anda \textit{et al.}~reported multistate RASPT2/RASSCF excitation energies of several BCL units within the light-harvesting system 2 of a purple bacterium \cite{Anda2016, Anda2019}. A RASPT2 approach was also combined with electrostatic embedding of fixed point charges to simulate the effect of the protein environment on excitation energies of the same system by Segatta \textit{et al.}~\cite{Segatta2017}. While these reports constitute important advances in the use of wavefunction-based methods for complex biological molecules, they were performed with relatively small basis sets and show a dependence on the choice of the restricted active space (RAS).

The \textit{ab initio} Bethe-Salpeter equation (BSE) approach, when rigorously based on many-body Green's function theory, is an alternative method for describing neutral excitations of correlated many-electron systems\cite{Onida2002}. It is based on a framework of charged excitation energies that correspond to electron addition and removal energies, and that are most frequently calculated within the $GW$ approximation. The $GW$+BSE approach has been shown to be successful in predicting the optical spectra of bulk solids \cite{Albrecht1998, Rohlfing2000}. and low-dimensional materials\cite{Qiu2013}. In recent years it has also begun to be applied to finite systems, such as small molecules \cite{Grossman2001,Tiago2008,Bruneval2015}, and larger molecular complexes \cite{Palummo2009,Duchemin2012,Wehner2018}, for which its accuracy has been shown to be comparable to single-reference wavefunction methods for both localized and charge transfer excitations \cite{Blase2018}, at substantially reduced computational cost.

In this article, we assess the accuracy of the \textit{ab initio} $GW$+BSE approach for the $Q_y$, $Q_x$ and the first bright $B$ excitation of several members of the BCL and CL family, and the chemically closely related Bacteriochlorin (BC) molecule. We compare two different approaches for approximating the electronic self-energy $\Sigma=iGW$: 1.~$G_0W_0$, a one-shot method, in which the zeroth order Green's function $G_0$ and screened Coulomb interaction $W_0$ are constructed from a DFT eigensystem and directly used to correct DFT eigenvalues perturbatively, 2.~partially self-consistent $GW$ (ev$G_nW_n$), in which the corrected eigenvalues are used to iteratively re-calculate $G$ and/or $W$ until self-consistency is reached. We compare our results to TDDFT calculations with the local density approximation (LDA), two global hybrid and an OT-RSH functional, with RASPT2 literature results \cite{Anda2016,Anda2019} and with experimental data \cite{Limantara1997,Scheer1978}.

We find that the $GW$+BSE approach used in a partially self-consistent manner, results in excitation energies in the visible and near-ultraviolet within less than 100\,meV from experiment for the entire family of pigments studied here. Our results are almost completely independent of the DFT eigensystem used as input for the $GW$+BSE calculations. In fact, even a simple and computationally inexpensive LDA starting point leads to excellent agreement with experiment and eliminates spurious charge transfer excitations between $Q_y$ and $Q_x$ that TDDFT with (semi)local functionals produces. Contrary to TDDFT, $GW$+BSE also correctly predicts the energy difference between the two $Q$-band excitations, a crucial prerequisite for understanding the coupling of excitations in systems consisting of more than one pigment. Finally, we show that differences between $evG_nW_n$+BSE and state-of-the-art TDDFT calculations using an OT-RSH functional, can be explained almost entirely based on differences in how electron-hole interactions are described by the xc kernel of TDDFT and the BSE kernel, respectively. Eigenvalue differences as computed with $evG_nW_n$ and DFT with an OT-RSH functional are almost identical.

The remainder of this article is structured as follows: We start by briefly reviewing the $GW$+BSE approach, and report computational details and numerical convergence. We then show the effect of different DFT starting points and partial self-consistency on the excitation energies of the BC molecule. After this, we discuss our results for BCL~\textit{a}, \textit{b}, \textit{c}, \textit{d} and \textit{e}, and CL~\textit{a}, and \textit{b}, followed by a comparison with literature results and an in-depth discussion of differences between our $GW$+BSE and TDDFT results for BCL~\textit{a}.

\section{Methods}
\label{sec:methods}
\subsection{The GW+BSE approach}
In Green's function-based many-body perturbation theory, the calculation of charged excitations, corresponding to electron removal and addition energies, is based on knowledge of the exact interacting single-particle Green’s function $G$, that can in principle be computed from a set of self-consistent integro-differential equations -- Hedin's equations -- linking $G$ to the electronic self energy $\varSigma$, the screened Coulomb interaction $W$, the irreducible polarizability $\chi$, and the vertex function $\Gamma$\cite{Fetter1971}. The lowest-order expansion of $\Sigma$ with respect to $W$, leads to the $GW$ approximation, in which the electronic self-energy $\varSigma=iGW$ \cite{Hedin1999}. 
Quasiparticle (QP) eigenvalues can be obtained by solving
\begin{equation}\label{eq:QP2}
\begin{multlined}
\left[-\frac{\hslash^2}{2m}\nabla^2 + V_{\text{ion}}(\mathbf{r}) + V_{\text{H}}(\mathbf{r})\right]\varphi_{n}^{\text{QP}}(\mathbf r) + \\ \int d \mathbf{r'}\Sigma(\mathbf r, \mathbf r';\varepsilon_{n}^{\text{QP}})\varphi_{n}^{\text{QP}}(\mathbf r')  = \varepsilon_{n}^{\text{QP}}\varphi_{n}^{\text{QP}}(\mathbf r).
\end{multlined}
\end{equation}
Here, $V_{\text{ion}}$ is the ionic potential, $V_{\text{H}}$ is the Hartree potential, and $\varepsilon_{n}^{\text{QP}}$ and $\varphi_{n}^{\text{QP}}$ are QP energies and wavefunctions, respectively.

To avoid the high computational cost of a self-consistent solution of Equation~\ref{eq:QP2}, the $GW$ approach is commonly used within a one-shot scheme, in which $G_0$ and $W_0$ are constructed from a (generalized) Kohn-Sham (gKS) eigensystem obtained from a preceding DFT calculation. We use the notation $G_0W_0$@gKS to refer to $G_0W_0$ based on the gKS eigensystem ($\varphi^{gKS}_{n}$;$\varepsilon^{gKS}_{n}$) computed with the xc functional $E_{\text{xc}}^{\text{gKS}}$. In this approach, QP corrections are calculated to first order in $\Sigma$ as
\begin{equation}
\label{eq:QP}
\varepsilon^{QP}_{n} = \varepsilon^{gKS}_{n} + \langle \varphi^{gKS}_{n} | \Sigma(\varepsilon_{n}^{\text{QP}}) - V_{xc} | \varphi^{gKS}_{n} \rangle,
\end{equation}
where $V_{xc}$ is the xc potential, and it is assumed that $\varphi^{QP}_{n} \approx \varphi^{gKS}_{n}$.

While the $G_0W_0$ approach has been used with much success, in particular for the calculation of band gaps and band structures of solids, a well-known and well-documented dependence on the gKS eigensystem used to construct $G_0$ and $W_0$, limits its predictive power \cite{Jiang2010,Liao2011,Marom2012}. Partial self-consistency in the QP eigenvalues can often mitigate this problem. In eigenvalue self-consistent $GW$, the gKS eigenvalues used to construct $G$ and/or $W$ are replaced with those from the output of a prior $GW$ step; the self-energy corrections are then iterated until the QP eigenvalues converge. This approach, that we call $evG_nW_n$ in the following ($n$ refers to the number of iterations), has been shown to remove much of the starting point dependence for a range of different systems \cite{Jacquemin2015,Kaplan2015}.

The BSE is an equation for the two-particle electron-hole Green’s function, and allows for the calculation of the polarizability including electron-hole interactions through the screened Coulomb interaction $W$. In practice, the BSE is usually solved neglecting the frequency dependence of $W$. Within this static approximation, it can be written in a form equivalent to Casida's equations of TDDFT
\begin{equation}
\label{eq:casida}
\begin{pmatrix} A & B \\ -B & -A \end{pmatrix} \begin{pmatrix} X^{s} \\ Y^{s} \end{pmatrix} = \Omega_{s} \begin{pmatrix} X^{s} \\ Y^{s} \end{pmatrix},
\end{equation}
where $\varOmega_s$ are neutral excitations and ($X^s$ , $Y^s$) are the corresponding eigenvectors \cite{Onida2002}. $A$ and -$A$ represent resonant and antiresonant transitions that can be expressed as
\begin{equation}
\label{eq:resonant}
A^{jb}_{ia} = (\varepsilon^{QP}_{a} - \varepsilon^{QP}_{i}) \delta_{ij} \delta_{ab} - 2(ia|jb) + W^{ab}_{ij}(\omega=0),
\end{equation}
and that are coupled through $B$ and -$B$, defined as
\begin{equation}
B^{jb}_{ia} = -2(ia|bj) + W^{ab}_{ij}(\omega=0),
\end{equation}
for singlet excitations. In these expressions $i$ and $j$ are occupied, and $a$ and $b$ are unoccupied states, and $(ia|bj)$ stands for
\begin{equation}
    (ia|bj) = \int\int d\mathbf rd\mathbf r' \varphi_i^{QP}(\mathbf r) \varphi_a^{QP}(\mathbf r) \frac{1}{|\mathbf r - \mathbf r'|} \varphi_j^{QP}(\mathbf r') \varphi_b^{QP}(\mathbf r'). 
\end{equation}
Note that, $\varphi_i^{QP}=\varphi_i^{gKS}$, whenever the $G_0W_0$ or $evG_nW_n$ approaches are used to construct $A$ and $B$.

\subsection{Computational Details}
Our calculations of charged and neutral excitations were performed using the $GW$+BSE and TDDFT implementation in the open source \textsc{molgw} software package (version 2B), which relies on Gaussian basis functions \cite{Bruneval2016}. We used the frozen-core approximation throughout, which changes excitation energies by less than 1\,meV. We also employed the resolution-of-the-identity (RI) method, in order to reduce the calculation of 4-center integrals to 2- and 3-center integrals. For BCL~\textit{a}, the RI changes the QP HOMO-LUMO gap by less than 50\,meV using a 6-31G basis set and BHLYP as a starting point, but we expect the effect of the RI to be even smaller for the larger basis sets used in the remainder of this article\cite{VanSetten2015}. 
To further reduce the computational cost of the evaluation of the $GW$ polarizability, we use the Single Pole Approximation (see Supporting Information for details). The Tamm-Dancoff approximation, which corresponds to neglecting the $B$ matrix elements in equation \ref{eq:casida} is not used as it consistently increases both $GW$+BSE and TDDFT results by $\sim$ 0.3\,eV, in agreement with previous findings\cite{Shao2020,Duchemin2012}.

We tested the influence of the Gaussian basis set size on HOMO-LUMO gaps and $Q_y$ and $Q_x$ excitations of BCL~$a$  (using a structure from Ref.~\citenum{Anda2016}) with $G_0W_0$@BHLYP+BSE, considering seven different basis sets, namely the Pople basis sets 6-31G, 6-311G, 6-311++G** and 6-311++G(2d,2p), combined with the DeMon auxiliary basis set \cite{Godbout1992}, and the Karlsruhe basis sets def2-SVP, def2-TZVP and def2-TZVPP and their corresponding auxiliary basis sets\cite{Zheng2011}.
%Figure1
\begin{figure}[htb]
\centering
\includegraphics[width=8cm]{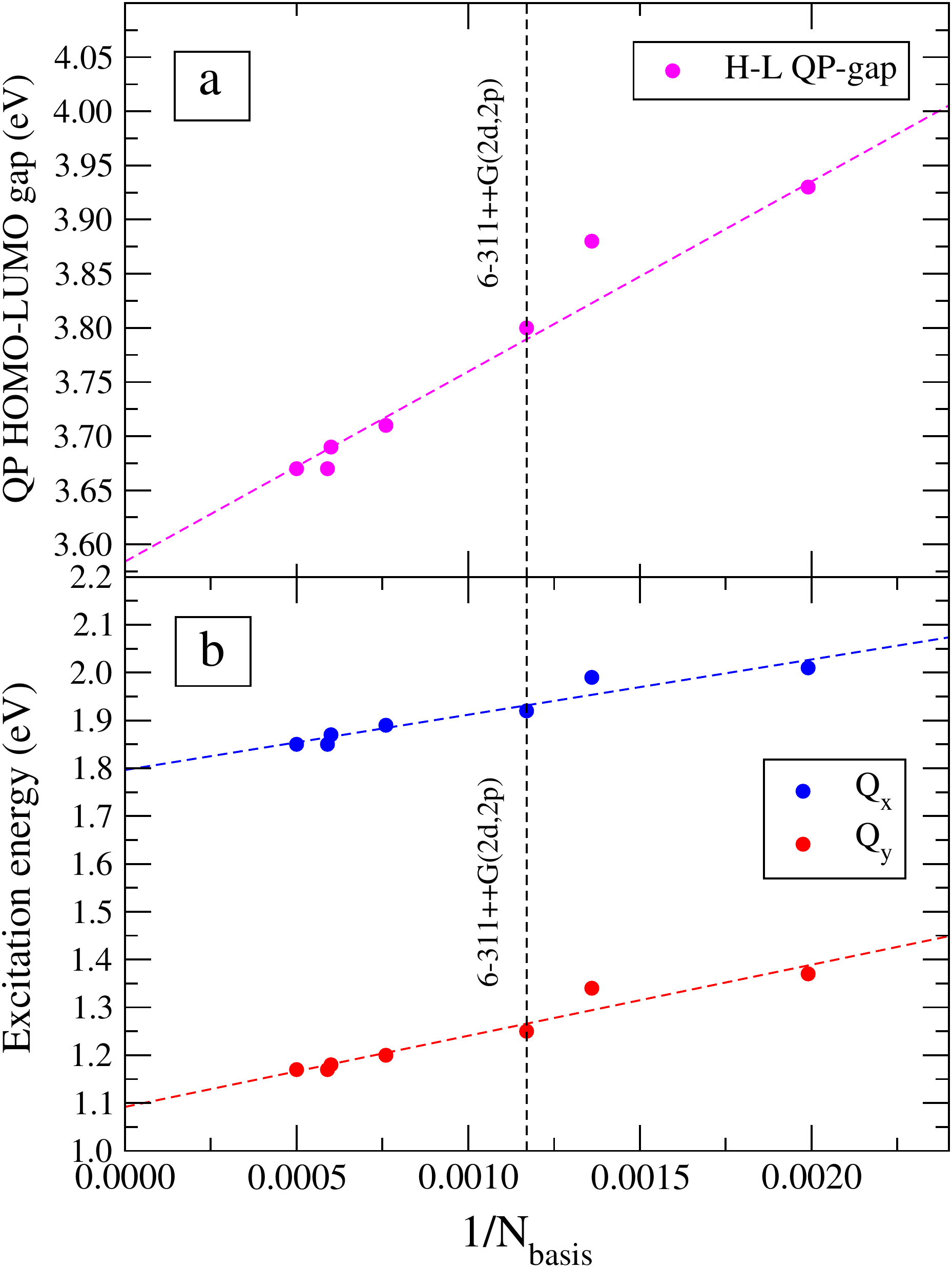}
\caption{\label{fig1}
Convergence as a function of number of basis functions 1/$N_{\text{basis}}$ for a) the HOMO-LUMO gap and b)  the $Q_y$ and $Q_x$ excitation energies, calculated with $G_0W_0$@BHLYP+BSE. Dashed lines represent a linear fit.}
\end{figure}

Figure~\ref{fig1} and Figure~S2 show the convergence of the HOMO-LUMO gap and $Q_y$ and $Q_x$ excitation energies as a function of the inverse number of basis functions, 1/$N_{\text{basis}}$ for $GW$+BSE and TDDFT, respectively (raw data is presented in Table S1 and S2). We find that the HOMO-LUMO gap depends significantly more on 1/$N_{\text{basis}}$ than $Q_y$ and $Q_x$ excitation energies, and that TDDFT results are less sensitive to the choice of basis set than $GW$+BSE. Based on these tests, we use the 6-311++G(2d,2p) basis set for all calculations reported in the following. We estimate the error in the $GW$(+BSE) HOMO-LUMO gap, and the $Q_y$ and $Q_x$ excitation energies by linearly extrapolating to an infinite basis set. By excluding the very small 6-31G and 6-311G basis sets from these fits, we obtain extrapolated values of 3.57\,eV for the HOMO-LUMO gap, 1.11\,eV for $Q_y$ and 1.81\,eV for $Q_x$, respectively. We conclude that by using the 6-311++G(2d,2p) basis set for all further calculations, we likely overestimate $GW$(+BSE) HOMO-LUMO gaps and $Q_y$ and $Q_x$ excitation energies by $\sim$0.1\,eV with respect to the complete basis set limit. Conversely, use of the Single Pole Approximation leads to a similar underestimation of the HOMO-LUMO gap, and the $Q_y$ and $Q_x$ excitations, resulting in a fortuitous cancellation of errors.

We test the effect of different xc functionals on our TDDFT and $GW$+BSE results. We use the LDA, two global hybrid functionals (B3LYP and BHLYP), and the range separated hybrid (RSH) functional $\omega$PBE. In RSH functionals, the Coulomb repulsion is separated into a short- and a long-range part, for numerical convenience expressed as
\begin{equation}
    \frac{1}{r} = \frac{1 - \text{erf}(\omega r)}{r} + \frac{\text{erf}(\omega r)}{r},
\end{equation}
where $\omega$ is called the range separation parameter. The $\omega$PBE functional uses PBE exchange in the short range and the exact exchange energy in the long range, allowing for a self-interaction free description at large electron-electron distances. We obtain the range separation parameter $\omega$ through the tuning procedure outlined in Ref.~\citenum{Stein2010}, in which $\omega$ is chosen such that the HOMO eigenvalue is as close as possible to the negative ionization potential both for the neutral and the anionic system. Consequently, and by construction, the resulting HOMO-LUMO gap is a very good approximation to the fundamental gap of the neutral molecule. We use the \textsc{qchem} code and a 6-31G(d,p) basis set for the tuning \cite{Shao2015}. The tuned range separation parameters for all systems discussed in the following, can be found in Table~S3.

\section{Results and Discussion}
\label{sec:results}

\subsection{Bacteriochlorin}
\label{sec:Bacteriochlorin}
To validate our methodological setup and investigate the starting point dependence of the $G_0W_0$ approach and the effect of eigenvalue self-consistency on our calculated HOMO-LUMO gaps and excitation energies, we start by examining the Bacteriochlorin (BC) molecule, for which $GW$+BSE results have been reported in Ref.~\citenum{Duchemin2012}. We use a BC structure from Ref.~\citenum{Duchemin2012} and denote the lowest energy excitations $Q_x$ and $Q_y$, respectively, according to the direction of their transition dipole moments. Table~\ref{table1} contains our calculated HOMO-LUMO gaps, $Q_x$ and $Q_y$ excitation energies and oscillator strengths using TDDFT and several flavors of the GW+BSE approach. We find that, as expected, (generalized) Kohn-Sham HOMO-LUMO gaps show a large dependence on the xc functional, with LDA, B3LYP and BHLYP leading to significantly lower and $\omega$PBE to a HOMO-LUMO gap similar to the HOMO-LUMO gap calculated with $G_0W_0$ and eigenvalue-self-consistent ev$G_nW_n$. In turn, $Q_y$ and $Q_x$ excitation energies from TDDFT are considerably less dependent on the xc functional than HOMO-LUMO gaps. In agreement with previous studies, we find that TDDFT overestimates the experimental values for $Q_x$ and $Q_y$ by up to $\sim$0.4\,eV, depending on the xc functional \cite{Duchemin2012}. TDDFT with the OT-RSH $\omega$PBE is in best agreement with experiment, overestimating it by $\sim$0.2\,eV for both excitations. We further find that $G_0W_0$@LDA+BSE underestimates $Q_x$ by 0.4\,eV and $Q_y$ by 0.6\,eV, whereas the use of a BHLYP and $\omega$PBE starting point results in excitations within 0.1\,eV of the experimental results.  In accordance with prior studies, we observe that most of the starting point dependence of the $G_0W_0$+BSE results is inherited from the starting point dependence of the HOMO-LUMO gaps \cite{Bruneval2015}.

In order to investigate the effect of eigenvalue self-consistency in the $GW$+BSE approach, we tested the effect of updating the eigenvalues in the construction of $G$ only ($evG_nW_0$), and of both $G$ and $W$ ($evG_nW_n$). Eigenvalue self-consistency in $G$ alone only slightly changes the results as compared to $G_0W_0$. In contrast, full eigenvalue self-consistency largely eliminates the starting point dependence. In particular, using an LDA starting point results in excitation energies within 0.1\,eV from experiment - similar to the $\omega$PBE starting point, but at considerably reduced computational cost. In Table S4, we report similar results for the more complex pigment BCL~$a$. In the remainder of this article we therefore focus primarily on eigenvalue self-consistent results based on LDA and $\omega$PBE starting points.
%Table1
\begin{table}[htb]
\centering
\begin{tabular}[t]{l|c|c|c|c|c|c}
\toprule
Method      &  xc functional    & H-L gap  &  Q$_x$  &   $\Gamma_x$      &  Q$_y$ & $\Gamma_y$ \\\midrule
\multirow{3}{*}{TDDFT} & LDA &     1.38      &  2.04   &    0.18        &  2.39  &     0.03     \\
& B3LYP                 &     2.17      &  2.06   &    0.23        &  2.51  &     0.04  \\
& BHLYP                 &     3.27      &  1.93   &    0.28        &  2.55  &     0.04 \\
& $\omega$PBE           &     4.38      &  1.87   &    0.23        &  2.42  &     0.05     \\\midrule
\multirow{3}{*}{$G_0W_0$+BSE} & LDA           &     4.15      &  1.21   &    0.09        &  1.67  &     0.04  \\
& B3LYP         &     4.36      &  1.44   &   0.14         &  1.97  &     0.04 \\
& BHLYP         &     4.56      &  1.67   &    0.19        &  2.23  &     0.05  \\
& $\omega$PBE   &     4.59      &  1.64   &    0.19        &  2.26  &     0.05   \\\midrule
\multirow{3}{*}{ev$G_nW_0$+BSE} & LDA         &     4.31      &  1.41   &  0.13      &  1.97  &   0.04 \\
& B3LYP       &     4.43      &  1.54   &    0.16        &  2.12  &    0.05     \\
& BHLYP       &     4.56      &  1.66   &    0.19        &  2.24  &   0.05    \\
& $\omega$PBE &     4.57      &  1.62   &    0.18        &  2.27  &   0.04  \\\midrule
\multirow{3}{*}{ev$G_nW_n$+BSE} & LDA         &     4.42      &  1.51   &    0.17        &  2.21  &     0.05  \\
& B3LYP       &     4.55      &  1.63   &    0.19        &  2.24  &    0.05    \\
& BHLYP       &     4.60      &  1.69   &    0.20        &  2.27  &     0.05       \\
& $\omega$PBE &     4.56      &  1.61   &    0.18        &  2.26  &     0.04    \\\midrule
Exp$^a$  &           &     ---       &  1.60   &    ---          &  2.30  &      ---    \\
\bottomrule
\end{tabular}
\caption{HOMO-LUMO gaps, $Q_x$ and $Q_y$ excitation energies (in eV) and corresponding oscillator strengths, $\Gamma_x$ and $\Gamma_y$, for BC calculated with the 6-311++G(2d,2p) basis set. $^a$ Data for bacteriopheophorbide from Ref.~\citenum{Scheer1978} and \citenum{Duchemin2012}.}
\label{table1}
\end{table}

\subsection{Excitation energies of Bacteriochlorophylls and Chlorophylls}
\label{sec:OtherPigments}
Next, we turn to reporting the vertical excitation energies of several members of the BCL and CL family of pigments. All structures were obtained from Ref.~\citenum{Oviedo2010} and geometry-optimized using DFT as implemented in the \textsc{Turbomole} code with a def2-TZVP basis set and the B3LYP xc functional \cite{Turbomole74}. Atomic coordinates of all relaxed structures can be found in the Supporting Information. We used both LDA and $\omega$PBE starting points for our ev$G_nW_n$+BSE, and $\omega$PBE for our TDDFT calculations. Unlike the $Q_x$ excitation of BCL~\textit{a} and \textit{b}, which has significant oscillator strength, the $Q_x$ excitation of BCL~\textit{c} -- \textit{e} is dark. Following Ref.~\citenum{Vokacova2007}, we therefore also compare our calculations with experimental results for the higher-energy $B$ band \cite{Scheer1978}. We report the vertical excitation energies and corresponding oscillator strengths of the first six excitations of all pigments in Table S6 and S7. In these calculations, we included a total of 20 excitations, in order to ensure that the higher lying excitations are well-converged.
%Table2
\begin{table*} [htb]
\centering
\begin{tabular}[t]{l||ccc|ccc|ccc|ccc}
\toprule
          &  \multicolumn{3}{c}{$GW$@LDA+BSE} &  \multicolumn{3}{c}{$GW$@$\omega$PBE+BSE}  & \multicolumn{3}{c}{TD-$\omega$PBE}   &   \multicolumn{3}{c}{Exp$^b$}    \\
 Molecule & Q$_y$ & Q$_x$ & B & Q$_y$ & Q$_x$ & B & Q$_{y}$ & Q$_{x}$ & B & Q$_{y}$ & Q$_{x}$ & B \\\midrule     
   BCL~\textit{a} & 1.52 & 2.08 & 3.25  & 1.50 & 2.10 & 3.16 & 1.75 & 2.16 & 3.33  & 1.60 & 2.15 & 3.46 \\
   BCL~\textit{b} & 1.48 & 2.07 & 2.95  & 1.45 & 2.09 & 3.05 & 1.69 & 2.15 & 3.19  & 1.56 & 2.14 & 3.37 \\
   BCL~\textit{c} & 1.85 & 2.05 & 2.94  & 1.84 & 2.11 & 3.02 & 2.05 & 2.21 & 3.21  & 1.88 &  --- & 2.89\\
   BCL~\textit{d} & 1.90 & 2.15 & 2.94  & 1.89 & 2.21 & 3.05 & 2.08 & 2.29 & 3.19  & 1.90 &  --- & 2.93\\
   BCL~\textit{e} & 2.01 & 2.04 & 2.78  & 1.96 & 2.13 & 2.88 & 2.10 & 2.23 & 3.02 & 1.92 &  --- & 2.72\\
   CL~\textit{a}  & 1.85 & 2.13 & 2.91  & 1.86 & 2.19 & 3.02  & 2.06 & 2.29 & 3.16 & 1.87 & 2.14 & 2.88\\
   CL~\textit{b}  & 1.95 & 2.17 & 2.79  & 1.93 & 2.20 & 2.85  & 2.10 & 2.29 & 2.97 & 1.92 & 2.26 & 2.72\\ \midrule
   MAE            & 0.05 & 0.06 & 0.12  & 0.04 & 0.05 & 0.17 & 0.17 & 0.05 & 0.25 & & & \\
  \bottomrule 
	\end{tabular}
    \caption{$Q_y$, $Q_x$, and first B band excitation energy of BCLs and CLs calculated using a 6-311++G(2d,2p) basis set. $GW$+BSE results are based on eigenvalue self-consistent ev$G_nW_n$. Experimental results in diethyl ether from Ref.~\citenum{Vokacova2007}.}
\label{table2}
\end{table*}

Table~\ref{table2} demonstrates that ev$G_nW_n$+BSE is in excellent agreement with experiment for the entire family of BCL and CL molecules. The MAE is about 50\,meV for the $Q_y$ and $Q_x$ and between 100 and 200\,meV for the $B$ excitation. Our ev$G_nW_n$+BSE results also accurately reflect the spectral shifts of the $Q_y$ excitation when comparing different BCL pigments with each other. For example, the BCL~\textit{b} molecule differs from BCL~\textit{a} through an ethyliden side group, which shifts the Q$_y$ excitation by 40\,meV to the red. This redshift is perfectly reproduced in our $GW$+BSE calculations. This is the first main result of this study. The second one is that our results are essentially independent of the DFT eigensystem used as input for the $GW$+BSE approach: A computationally inexpensive LDA starting point results in the same level of agreement with experiment as the more tedious $\omega$PBE calculation that involves a system-dependent tuning procedure for the range separation parameter $\omega$. This is in stark contrast to TDDFT. TD-LDA leads to spurious excitations with charge transfer character in between $Q_y$ and $Q_x$, as well as slightly above Q$_x$, depending on structure, as discussed below and in the literature\cite{Schelter2019}. TDDFT with the optimally-tuned $\omega$PBE results in good agreement with experiment for all three excitations, albeit with slightly higher MAEs of 170\,meV, 50\,meV, and 250\,meV for $Q_y$, $Q_x$ and $B$, respectively.

\begin{figure}
\centering
\includegraphics[width=\columnwidth]{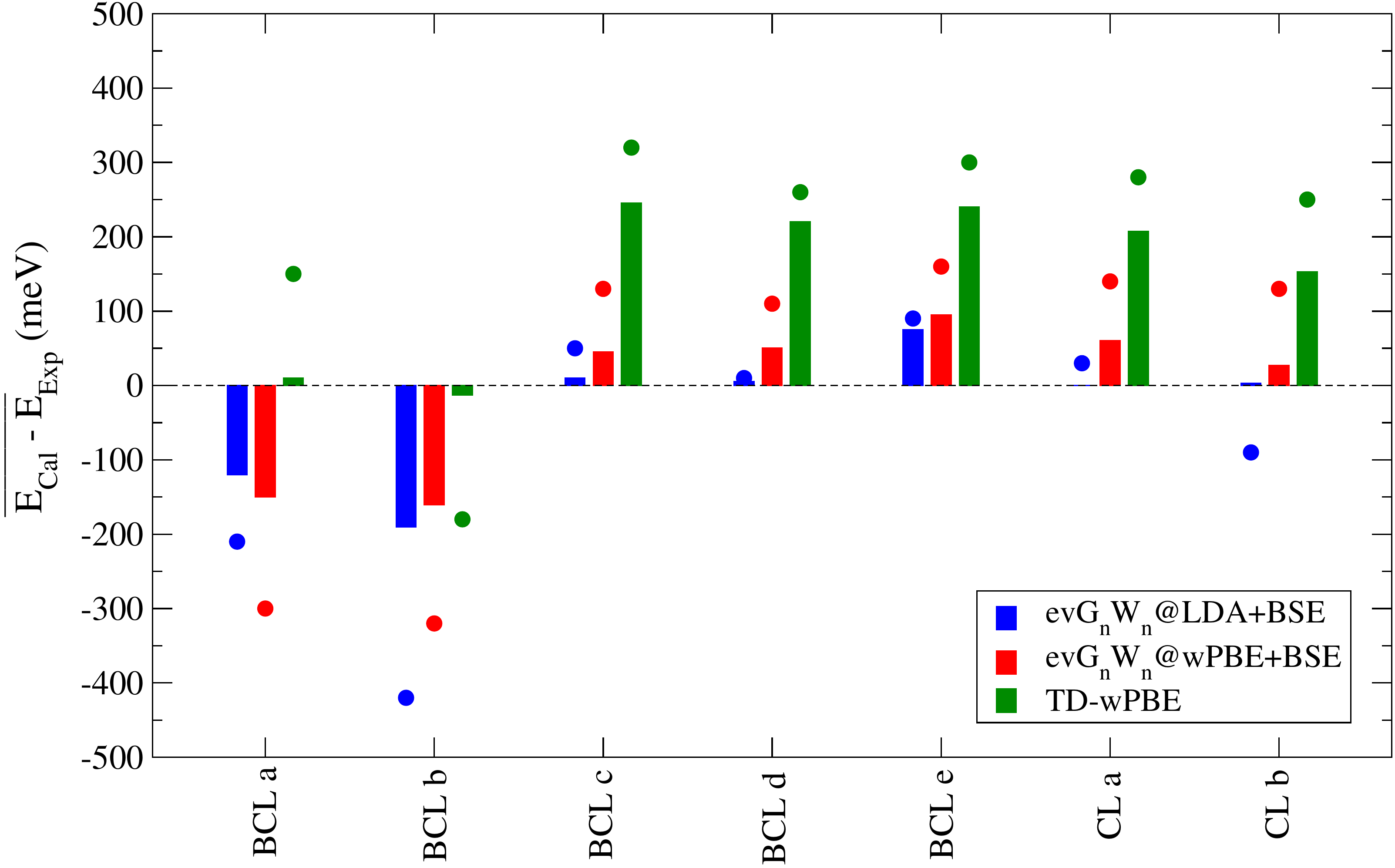}
\caption{\label{fig2} Colored bars denote the average difference between calculated and experimental excitation energies for ev$G_nW_n$@LDA+BSE (blue), ev$G_nW_n$@$\omega$PBE+BSE (red) and TDDFT (green) with $\omega$PBE. The dots represent the maximum deviation in each case.}
\end{figure}

In Figure~\ref{fig2} we plot the difference between our calculated results and experiment, averaged over all three excitations, to further highlight qualitative differences between ev$G_nW_n$+BSE and TDDFT. For BCL~$a$ and BCL~$b$, ev$G_nW_n$+BSE on average underestimates experiment by $\sim$100\,meV, whereas the average TDDFT deviation is close to zero, because TDDFT slightly overestimates the $Q_y$ and $Q_x$ excitations, but underestimates the B excitation of these pigments. For all other BCL and the two CL molecules studied here, we consistently find that the average deviation of ev$G_nW_n$+BSE is significantly smaller than that of TDDFT. Similar to our results for the BC molecule and to other benchmark studies of complex organic molecules \cite{Shao2020}, TDDFT tends to overestimate all three excitations by between 200 and 300\,meV. ev$G_nW_n$+BSE is in much closer agreement with experiment for these pigments, on average overestimating their excitation energies by less than 100\,meV. We stress again, that these results are independent of the DFT starting point, whereas our TDDFT results rely on a per-system tuning procedure.

Our results are in excellent agreement with correlated excited states methods for those systems for which such studies have been reported, primarily CL~$a$ and BCL$~a$. ADC(2) excitation energies of the first three excitations of CL~$a$ reported by Suomivuori \textit{et al.} are 1.85\,eV, 2.13\,eV, and 2.91\,eV, within $\sim$0.1\,eV of our ev$G_nW_n$@LDA+BSE results \cite{Suomivuori2019}. In another study by the same authors, the ADC(2) Q$_y$ excitation energy of histidin-ligated BCL~$a$ was reported to be 1.46\,eV, again within $\sim$0.1\,eV of our results, although it should be noted that the structures of ligated and free-standing BCL~$a$ slightly differ, leading to excitation energy differences of 10 - 30\,meV at the ADC(2) level \cite{Suomivuori2016}. Furthermore, Sirohiwal \textit{et al.} used a pair-natural orbital coupled cluster approach to study CL~$a$, and reported Q$_y$ and Q$_x$ excitation energies of 1.75\,eV and 2.24\,eV, respectively, for CL$~a$, also within $\sim$0.1\,eV of our GW+BSE results for these excitations \cite{Sirohiwal2020}.

%Figure2
%Table3
\begin{table}[ht]
\centering
\begin{tabular}[t]{l|c|c}
\toprule
Method      &  xc functional    & $\Delta_{Q_x - Q_y}$ \\\midrule
\multirow{4}{*}{ev$G_nW_n$+BSE}   & LDA           & 0.57 \\
                                    & B3LYP         & 0.54 \\
                                    & BHLYP         & 0.54 \\
                                    & $\omega$PBE   & 0.59  \\\midrule
\multirow{4}{*}{TDDFT}              & LDA           & 0.25  \\
                                    & B3LYP         & 0.40 \\
                                    & BHLYP         & 0.64    \\
                                    & $\omega$PBE   & 0.43  \\\midrule
Exp\cite{Vokacova2007}              &               & 0.55  \\
\bottomrule
	\end{tabular}
	\caption{Difference between $Q_x$ and $Q_y$ excitation energies (in eV) using TDDFT and ev$G_nW_n$+BSE for BCL~$a$.}
\label{table3}
\end{table}
Not only the absolute energies of $Q_y$ and $Q_x$ excitations are important for understanding and predicting excitation energy and charge transfer in photosynthetic systems, but also their relative energy difference, $\Delta_{Q_x - Q_y}$, plays a role, in particular for coupled systems of several pigment units. It is therefore reassuring that $evG_nW_n$+BSE predicts $\Delta_{Q_x - Q_y}$ in very good agreement with experiment, with a deviation of only 10\,meV for BCL~$a$, BCL~$b$ and CL~$a$, and 120\,meV for CL~$b$ for the LDA starting point, and a slightly larger deviation of on average 60\,meV for the $\omega$PBE starting point. TDDFT based on $\omega$PBE tends to underestimate $\Delta_{Q_x - Q_y}$, by on average 110\,meV for these four pigments. For BCL~$a$, we also show in Table~\ref{table3} that $\Delta_{Q_x - Q_y}$ strongly depends on the xc functional used in the TDDFT calculations, primarily because of the strong dependence of the $Q_x$ excitation on the amount of exact exchange, which can be seen by comparing the results based on the LDA (0\% of exact exchange), B3LYP ($\sim$ 23\%) and BHLYP (50\%). As before, ev$G_nW_n$+BSE is in excellent agreement with experiment, and almost independent of the underlying xc functional.

The experimental results reported in Table~\ref{table2} and \ref{table3} are based on measurements in diethyl ether, whereas our calculations are for gas-phase molecules. To approximately account for the effect of the solvent, we extracted experimental reference values for $Q_y$ and $Q_x$ excitations from a study by Limantara \textit{\textit{et al.}} \cite{Limantara1997}, in which electronic absorption spectroscopy was used to obtain $Q_y$ and $Q_x$ for a large number of nonpolar and polar solvents at room temperature. This study reports regression lines for $Q_y$ and $Q_x$ excitations of BCL~$a$ as a function of $R(n)=n^2 - 1/n^2 +2$, where $n$ is the refractive index of the solvent. The extrapolated values for $n=1$ (vacuum) are, 1.68\,eV (nonpolar) and 1.67\,eV (polar) for the $Q_y$ and 2.25\,eV (nonpolar) and 2.21\,eV (polar) for the $Q_x$ excitation. Based on these regression parameters, we estimate that the experimental reference values in Table~\ref{table2} lie $\sim$50-70\,meV below the gas phase excitation energies. We also calculated the Q$_y$ and Q$_x$ excitation energies of BCL~$a$ with TDDFT (using $\omega$PBE), approximating solvent effects with the COSMO approach as implemented in \textsc{turbomole}. We used a dielectric constant of 4.33\,$\epsilon_0$ corresponding to the value in diethyl ether. COSMO red-shifts the Q$_y$ and Q$_x$ excitation energies by 70\,meV and 50\,meV, respectively, supporting our estimate. We conclude that solvent effects are small - within the numerical accuracy of our $GW$+BSE calculations - and do not change our main conclusions. Note that we also neglect the effects of temperature and the 0-0 vibrational energy contribution in our comparison with experimental results. Exact agreement of our calculated results with experiment is therefore not expected.

\subsection{Bacteriochlorophyll \textit{a}}
\label{sec:BLCa}
In the remainder of this paper, we will use the BCL~$a$ molecule as a case study to compare to available computational literature results for this pigment, discuss the origin of differences between our $evG_nW_n$+BSE and TDDFT results, and comment on the effects of the choice of structure on excitation energies.
%Figure3
\begin{figure}
\centering
\includegraphics[width=0.6\columnwidth]{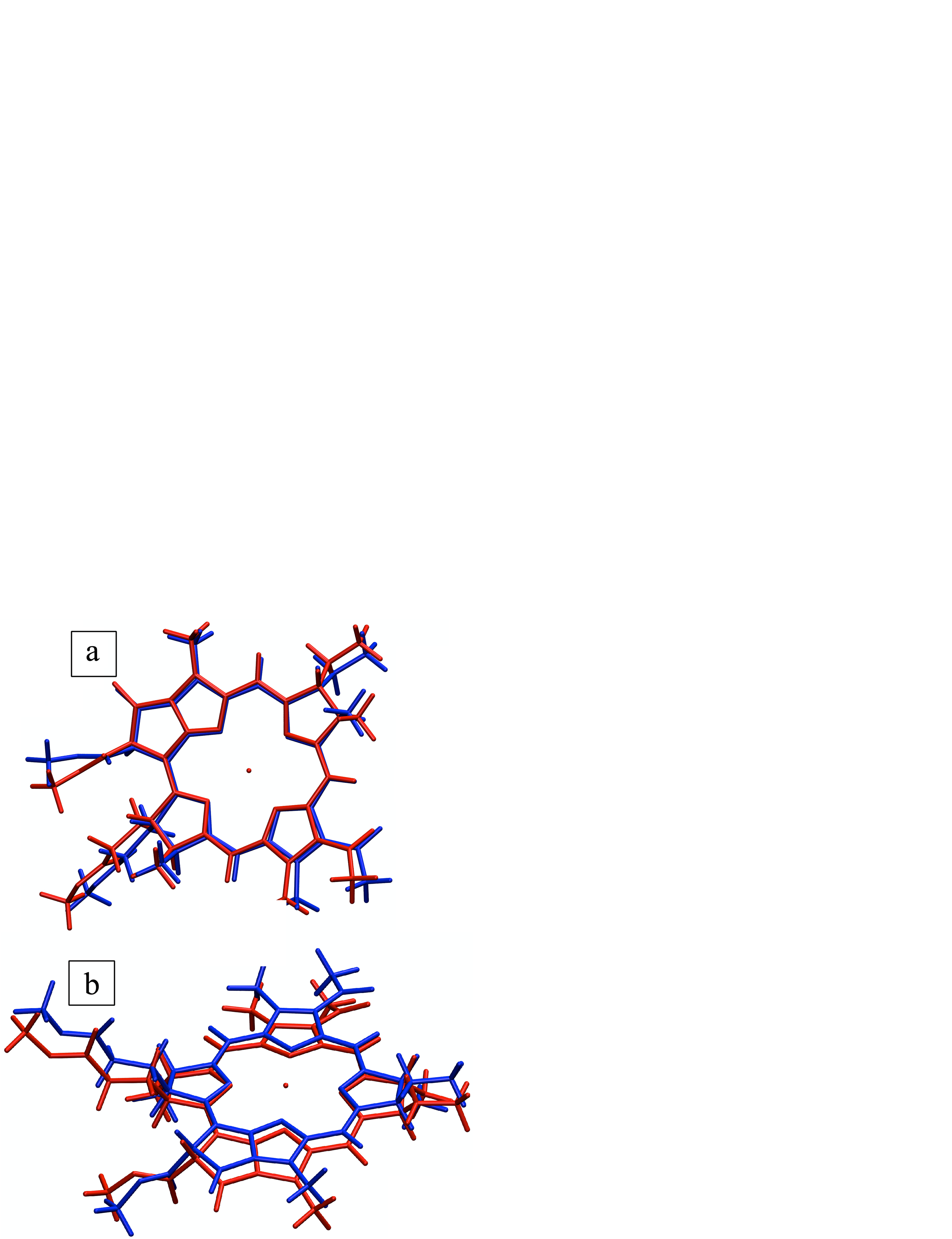}
\caption{\label{fig3} Overlay of structures 'A'(red) and 'R' (blue) in a) top view and b) side view.}
\end{figure}

\textit{\textbf{Comparison with RASPT2:}}~For the $Q_y$ and $Q_x$ excitations of BCL~$a$, we compare our $GW$+BSE and TDDFT calculations to multistate, second order perturbation theory (RASPT2) calculations by Anda \textit{et al.} \cite{Anda2016,Anda2019}. For this comparison, we use the molecular geometry reported in Ref.~\citenum{Anda2016}, which is a BCL~\textit{a} unit from the light-harvesting system LH2 of \textit{Rhodoblastus acidophilus}. This structure was extracted from an experimental X-ray crystallographic structure of the LH2 complex (unit 302 within structure 1NKZ in the RCSB Protein Data Bank) \cite{Papiz2003}. The phytyl tail was truncated and replaced by a hydrogen atom, and no further geometry optimization was carried out. In the following, we will call this structure 'A'. Our geometry-optimized version of 'A', which we relaxed using DFT as implemented in the \textsc{Turbomole} code with a def2-TZVP basis set and B3LYP\cite{Turbomole74} will be called 'R'. A visual comparison between 'A' and 'R' is shown in Figure~\ref{fig3}. The large differences that we observe between these two structures are unsurprising, given that we perform our geometry optimizations without taking into account the protein environment in which BCL~\textit{a} 'A' is embedded in \textit{in vivo}. Table~\ref{table4} shows our $GW$+BSE and TDDFT results for 'A' in comparison with the RASPT2 excitation energies from Ref.~\citenum{Anda2016} and \citenum{Anda2019}. We find, as before, that when eigenvalue self-consistency is used in $GW$, HOMO-LUMO gaps and $Q_y$ and $Q_x$ excitation energies differ by a maximum of 0.1\,eV. Most notably, however, our $GW$+BSE excitation energies substantially differ from those calculated with RASPT2, with $Q_y$ 0.4\,eV and $Q_x$ 0.5\,eV lower than the RASPT2 result.
%Table4
\begin{table}[ht]
\centering
\begin{tabular}[t]{c|c|c|c|c}
\toprule
Method      &  xc functional    & H-L gap  &  Q$_y$     &  Q$_x$ \\\midrule
\multirow{4}{*}{\parbox{4cm}{ev$G_nW_n$+BSE \\ 6-311++G(2d,2p) }} & LDA               &      3.62    &    1.17   &  1.90      \\
& B3LYP             &      3.67    &    1.19     &  1.90        \\
& BHLYP             &      3.72    &    1.23    &  1.92       \\
& $\omega$PBE       &      3.68    &    1.16   &  1.91      \\\midrule
\parbox{4cm}{ev$G_nW_n$+BSE \\ ANO-RCC-vDZP} & LDA & 3.76 & 1.38 & 2.18 \\ \midrule
\multirow{4}{*}{\parbox{4cm}{TDDFT \\ 6-311++G(2d,2p) }} & LDA                         &      0.92    &    1.59    &  1.99      \\
& B3LYP                       &      1.60    &    1.64    &  2.17      \\ 
& BHLYP                       &      2.61    &    1.57     &  2.34      \\
& $\omega$PBE                 &      3.70    &    1.48   &  2.02     \\ \midrule
\parbox{4cm}{RASPT2 \\ ANO-RCC-vDZP} &                      &              &    1.61     &  2.40  \\    \bottomrule
\end{tabular}
\caption{HOMO-LUMO gaps, $Q_y$ and $Q_x$ excitation energies (in eV) for BCL~\textit{a} structure 'A'.}
\label{table4}
\end{table}

We find that about half of this difference can be traced back to the use of a smaller basis set (ANO-RCC-vDZP) in Ref.~\citenum{Anda2016}. Repeating our ev$G_nW_n$@LDA+BSE calculation with the same basis, we obtain excitation energies of 1.38\,eV for Q$_y$ and 2.18\,eV for Q$_x$, respectively. In line with previous studies, we also find that TDDFT with global hybrid functionals (B3LYP and BHLYP) results in similar excitation energies as RASPT2 for the $Q_y$ excitation\cite{Anda2019,List2013}. We hypothesize that this agreement is fortuitous. The optimally-tuned RSH functional $\omega$PBE has been shown to better describe singlet excitation energies of a wide variety of organic compounds as compared to global hybrid functionals \cite{Kronik2012,Refaely-Abramson2011,Jacquemin2014}, and is more than 0.1\,eV lower in energy than the RASPT2 Q$_y$ excitation energy. Similar trends have also been shown for CL~$a$, where DFT-based multireference CI, just as TDDFT with global hybrid functionals, tends to overestimate experiment by $\sim$0.2\,eV for the $Q_y$ and the $Q_x$ excitation\cite{Parusel2000}. All in all, given that comparisons with experimental data are complicated for an \textit{in vivo} structure as 'A', we consider it most likely that our $GW$+BSE calculations underestimate the excitation energies of structure 'A' by $\sim$0.1\,eV, similar to our results for gas-phase BCL~$a$ (Table~\ref{table2}). The remaining deviations could be attributed to the multireference character of the Q$_y$ excitation\cite{Anda2016} and the choice of the restricted active space. 

We also note that our GW+BSE results reproduce the energetic order and relative energy differences of the Q$_y$ excitation of other BCL units within the LH2 ring that RASPT2 predicts, when using the ANO-RCC-vDZP basis set. However, use of the significantly larger 6-311++G(2d,2p) basis leads to substantially larger excitation energy differences between these units (Table S7). Finally, it is worth mentioning that our $GW$+BSE calculations reproduce the relatively large energy difference $\Delta_{Q_x-Q_y} \approx 0.8$\,eV that RASPT2 predicts, whereas TDDFT excitation energy differences are much less sensitive to details of the structure, with $\Delta_{Q_x-Q_y}\approx 0.5$\,eV (using $\omega$PBE) similar to the gas-phase structure of BCL~$a$. We speculate that a geometry optimization of structure 'A' within its protein environment would result in a smaller $\Delta_{Q_x-Q_y}$ for both RASPT2 and $GW$+BSE.
%Figure 4
\begin{figure}[htb]
\centering
\includegraphics[width=\columnwidth]{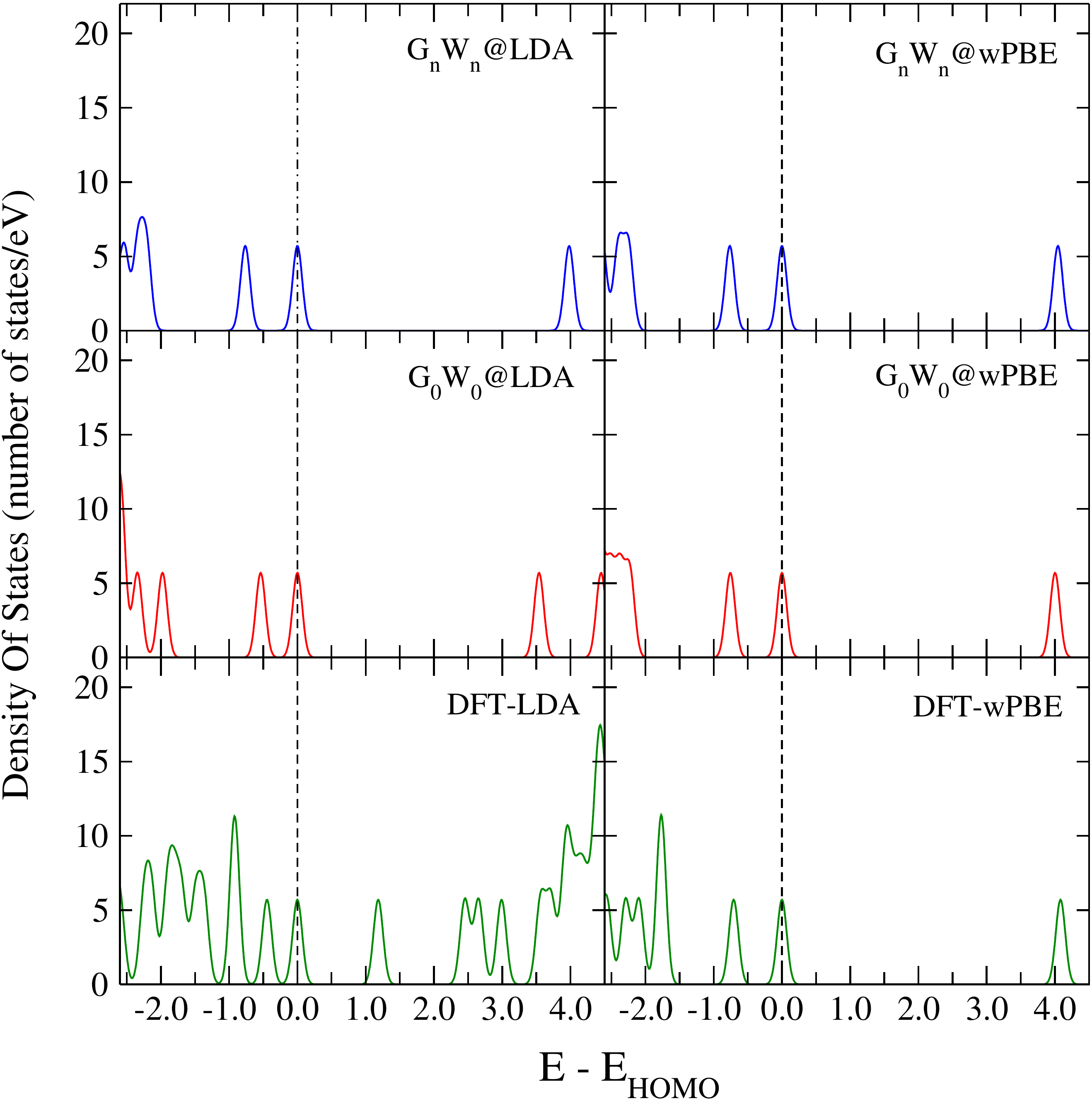}
\caption{\label{fig4}
(Generalized) Kohn-Sham (green), $G_0W_0$ (red) and $evG_nW_n$ (blue) DOS calculated using the LDA and $\omega$PBE. The HOMO energies are aligned to zero.}
\end{figure}

\textit{\textbf{Role of the electron-hole kernel:}}~We find that the difference between $GW$+BSE and TDDFT excitation energies can be traced back almost entirely to differences in how electron-hole interactions are described in both schemes. The $Q_y$ excitation is primarily ($\sim$90\%) a HOMO $\rightarrow$ LUMO transition, and the HOMO-LUMO gaps as calculated with DFT-$\omega$PBE and $evG_nW_n$@$\omega$PBE differ by only 0.02\,meV (Table~\ref{table3}). In fact, the density of states (DOS) in the energy range relevant for both the $Q_y$ and the $Q_x$ excitations based on $evG_nW_n$@$\omega$PBE and DFT-$\omega$PBE eigenvalues, are almost identical (see Figure~\ref{fig4}). To further test our hypothesis, we construct the statically screened Coulomb interaction $W_{ij}^{ab}$ (see Equation~\ref{eq:resonant}) and solve the BSE based on an DFT-$\omega$PBE eigensystem (instead of first computing QP eigenvalues using Equation~\ref{eq:QP}). We obtain values for the $Q_y$ and $Q_x$ excitation that are only 20\,meV higher and 40\,meV lower than the full $GW$+BSE solution, respectively, for structure 'A'. Similarly, for structure 'R', the results are within less than 10\,meV and 50\,meV for the $Q_y$ and $Q_x$ excitation, respectively. This observation confirms that differences between the $GW$+BSE and TDDFT excitation energies are primarily due to differences in the xc and the BSE kernel. Generally, the overestimation of excitation energies that we observe with TDDFT is in line with results for other organic $\pi$ chromophores such as rhodamine and rosamine\cite{Moore2017}, and of phenothiazine dyes \cite{DeQueiroz2021}, for which it has been linked to an insufficient treatment of differential electron correlation between the ground and excited states by most TDDFT xc kernels \cite{Moore2013}.

\begin{figure}
\centering
\includegraphics[width=\columnwidth]{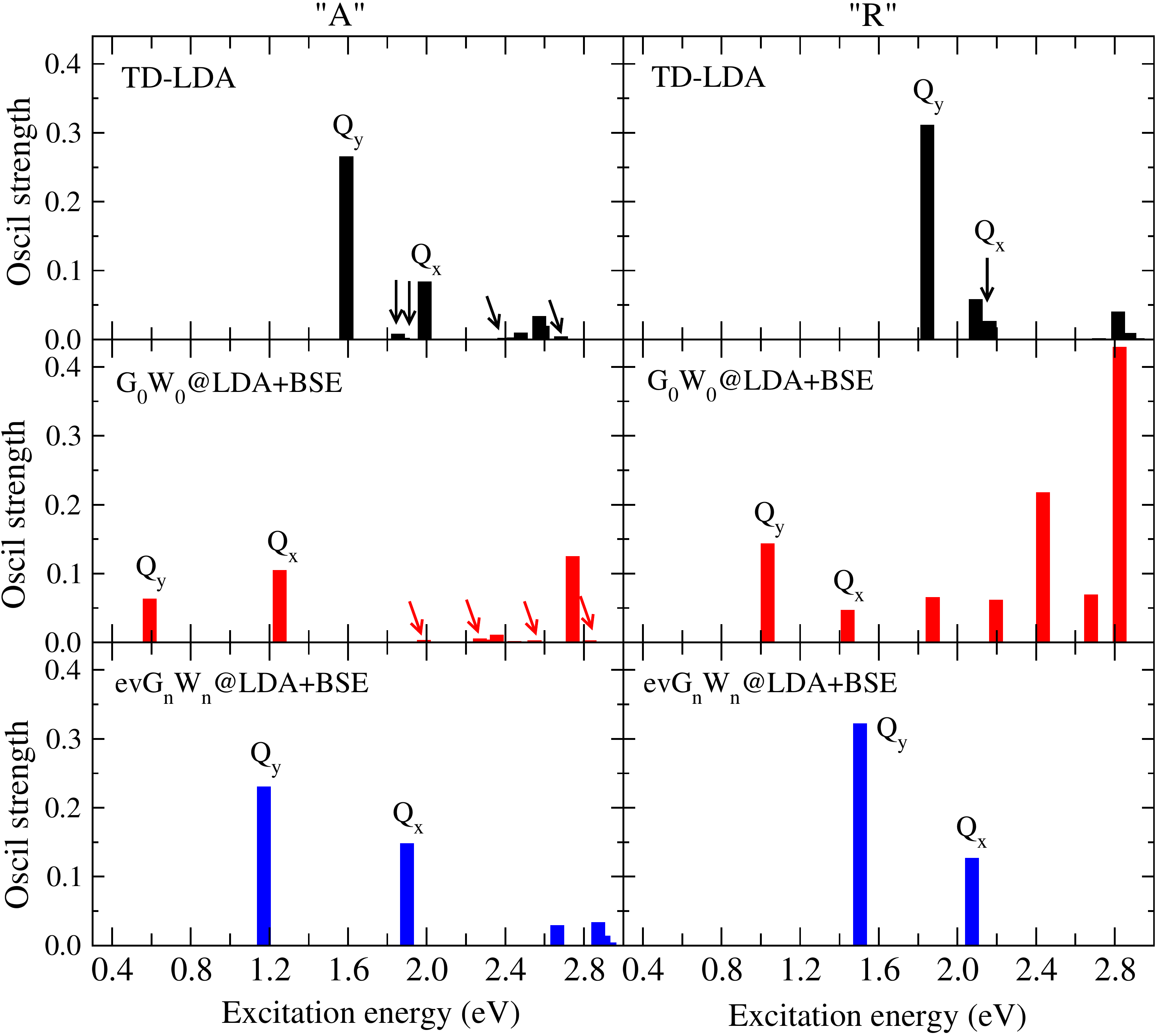}
\caption{\label{fig5}
First excitations for structures 'A' (left) and 'R' (right) as calculated with TD-LDA (top), $G_0W_0$@LDA+BSE (center) and ev$G_nW_n$@LDA+BSE (bottom). Arrows indicate excitations with very low oscillator strength.}
\end{figure}

\textit{\textbf{Charge transfer excitations with TD-LDA and $G_nW_n$@LDA+BSE:}}~Finally, motivated by the excellent performance of ev$G_nW_n$@LDA+BSE, we compare $G_0W_0$@LDA+BSE, ev$GW$LDA+BSE and TD-LDA results for structures 'A' and 'R' of BCL~$a$. Figure~\ref{fig5} shows the excitation spectrum calculated at these levels of theory. TD-LDA's severe underestimation of charge transfer excitations is well-known \cite{Dreuw2004} and leads to spurious excitations with charge transfer character at energies between $Q_y$ and $Q_x$ for BCL~\textit{a} \cite{Schelter2019}. Our comparison of structures 'A' and 'R' shows that while the energy of $Q_y$ and $Q_x$ is changing only slightly when TD-LDA is used, the relative position of these spurious low-oscillator strength excitations depends strongly on the structure. $G_0W_0$@LDA+BSE results in a very different, albeit no more reassuring picture. For both structures, the first excitation already appears at energies below or around 1\,eV and its oscillator strength is considerably lower than with TD-LDA; for structure 'A' the oscillator strength of $Q_y$ is even lower than that of $Q_x$. For structure 'R', excitations 2, 3 and 4 have similar, very low, oscillator strength. However, already at the $G_0W_0$@LDA+BSE level, no charge transfer excitations are found between Q$_y$ and Q$_x$ -- a consequence of the inherent non-locality of the BSE kernel. Finally, for both structures, eigenvalue self-consistency pushes all excitations to significantly higher energies and results in a quantitatively correct description of $Q_y$ and $Q_x$.

Inspection of the DOS calculated with DFT-xc, $G_0W_0$@xc and $G_nW_n$@xc (xc=LDA, $\omega$PBE) shown in Figure~\ref{fig4} is instructive for understanding the contribution of eigenvalue differences to the TDDFT and $GW$+BSE excitation energies. The $G_0W_0$@LDA DOS underestimates the HOMO-LUMO gap and the energy difference between the HOMO and HOMO-1. In contrast, there is virtually no difference between the HOMO, HOMO-1 and LUMO energies as calculated with DFT-$\omega$PBE, $G_0W_0$@$\omega$PBE and ev$G_nW_n$@$\omega$PBE, and ev$G_nW_n$@LDA. As expected, the DFT-LDA DOS is markedly different, underestimating the HOMO-LUMO gap, but also significantly underestimating the energy differences between the HOMO-1, HOMO-2 and HOMO-3. Notably, the spurious dark states between $Q_y$ and $Q_x$ that TD-LDA predicts, have significant contributions from transitions involving these lower occupied states.

\section{Conclusions}
In this article, we performed a systematic first principles study of the electronic structure and excitations of seven members of the (bacterio)chlorophyll family, which we validated through comparison with calculated and experimental literature results. The $GW$+BSE approach, when used in a partially self-consistent fashion, is in excellent agreement with experiment for excitations in the visible and near-ultraviolet part of the spectrum. $GW$+BSE also correctly predicts the energy difference between the low-energy $Q_y$ and $Q_x$ excitations of these pigments, relevant for the description of the coupling between pigment complexes, present in the light harvesting units and reaction centers of plants and bacteria and crucial for excitation energy and charge transfer. Most importantly, our results are almost entirely independent of the DFT eigensystem used as input for the $GW$+BSE calculations. A computationally inexpensive LDA starting point leads to similar results as a more involved optimally tuned $\omega$PBE starting point. 

It should be noted that the $GW$ approach, despite its implementation using Gaussian basis functions and the use of the RI approximation in \textsc{molgw} and other codes, remains a major bottleneck of these calculation due to its $O(N^4$) scaling with system size. Furthermore, our results highlight that the $GW$ approach, more so than DFT, requires careful convergence with respect to the basis set size. This limits its applicability to systems with a few (B)CL pigments at most, until algorithms with better scaling become more widely available \cite{Neuhauser2014, Vlcek2017,Foerster2020}. Our study joins a growing number of results demonstrating that the $GW$+BSE approach can accurately predict neutral excitations of complex molecules without empirical parameters \cite{Blase2018}. With new approaches for combining $GW$+BSE with large scale molecular mechanics simulations \cite{Wehner2018} and polarizable continuum embedding \cite{Duchemin2018} emerging, an accurate prediction of excitation energy and charge transfer in complex molecular environments is within reach.

%\begin{suppinfo}
%Additional convergence data, tuned range-separation parameters, %excitation energies of all molecules discussed in the main manuscript and %atomic coordinates.
%\end{suppinfo}

\begin{acknowledgments}
The authors are grateful for helpful discussions with C. Filippi. This work was supported by the Bavarian State Ministry of Science and the Arts through the Collaborative Research Network Solar Technologies go Hybrid (SolTech), the Elite Network Bavaria (ENB), and through computational resources provided by the Bavarian Polymer Institute (BPI).
\end{acknowledgments}

\end{document}